\documentclass[reprint,amsmath,amssymb,showpacs,aps,prl]{revtex4-1}
\usepackage{graphicx}
\usepackage{dcolumn}
\usepackage{bm}
\usepackage{float}
\usepackage{hyperref}
\usepackage{color}
\usepackage{cancel}

\newcommand{\ket}[1]{\left| #1\right\rangle}

\newcommand{\be}[0]{\begin{equation}}
\newcommand{\ee}[0]{\end{equation}}

\newcommand{\lra}\simeq
\newcommand{\eeqref}[1]{Eq.~(\ref{#1})}

\begin{document}
\preprint{APS/123-QED}
\title{Quantum vampire:\\ collapse-free action at a distance by the photon annihilation operator.}
\author{Ilya A. Fedorov$^{1,2}$}
\author{Alexander E. Ulanov$^{1,3}$}
\author{Yury V. Kurochkin$^1$}
\author{A. I. Lvovsky$^{1,2,4}$}\email{LVOV@ucalgary.ca}
\affiliation{$^1$Russian Quantum Center, 100 Novaya Street, Skolkovo, Moscow 143025, Russia}
\affiliation{$^2$P. N. Lebedev Physics Institute, Leninskiy prospect 53, Moscow 119991, Russia}
\affiliation{$^3$Moscow Institute of Physics and Technology, 141700 Dolgoprudny, Russia}
\affiliation{$^4$Institute for Quantum Science and Technology, University of Calgary, Calgary AB T2N 1N4, Canada}

\date{\today}
\begin{abstract}
A nonclassical state of light is distributed, via a beam splitter, between two remote parties. One of the parties applies the photon annihilation operator to its portion of the state. Surprisingly, this local intervention removes a photon from the entire initial state, leaving its mode as well as the spatial and temporal structure undisturbed. In this way, nonlocal quantum action-at-a-distance occurs without local state collapse by either party. This leads to curious consequences, such as the absence of a shadow when the annihilation operator is applied to a part of the spatial cross-section of the initial optical mode. In the experiment, we subtract a single photon from a part of an optical mode initially prepared in the one- or two-photon Fock state. Subsequent homodyne tomography reveals that the whole mode has jumped to the next lower Fock state, with no change in the mode shape.
\end{abstract}

\maketitle
One of the most intriguing and fundamental aspects of quantum mechanics is nonlocality. Discovered about 80 years ago, it became a basis for a lot of fundamental research and practical applications. To date, most studies of quantum action at a distance were based on local application of the projection measurement of the von Neumann type to an entangled state initially shared between two or more parties. On application of the projection operator, the state collapses, modifying the physical reality at a remote location in a nonlocal fashion. 

Here we implement action at a distance with a local operation of a different type --- the photon annihilation operator, applied by one of the remote parties to its portion of a shared optical state. Unlike von Neumann measurements, the annihilation operator does not collapse the entangled state, but only modifies it. One would intuitively expect this modification to be of local nature, having effect only on the optical mode to which it is applied. However, as we find both theoretically and experimentally, the action of the annihilation is sometimes global, removing the photon from the entire entangled state.

\begin{figure}[t]
\includegraphics[width=3.4in]{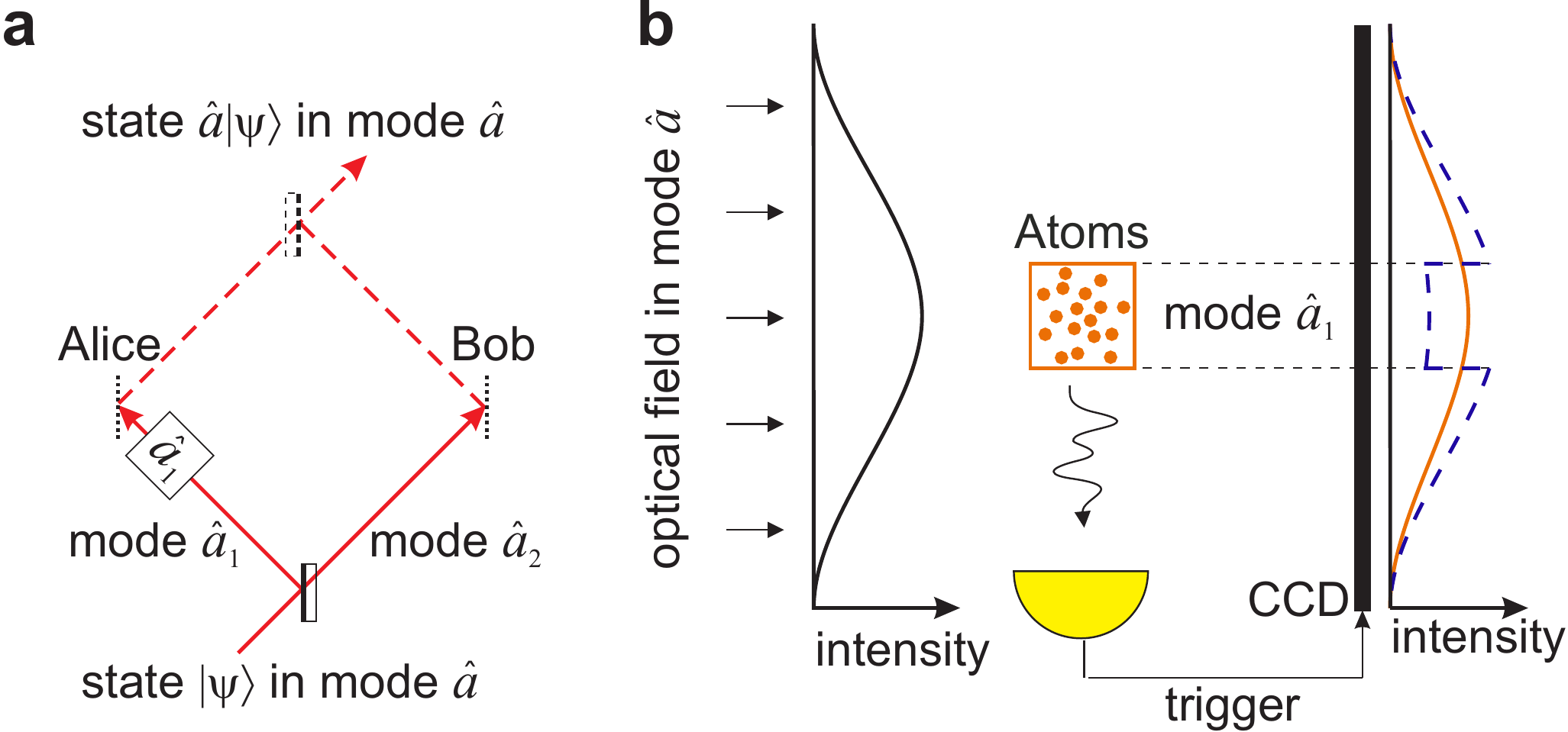}
\caption{The quantum vampire effect. a) When state $\ket\psi$ in the mode defined by annihilation operator $\hat a$ is split between two remote parties, the application of the photon annihilation operator $\hat a_1$ by one of these parties affects state $\ket\psi$ globally. This can be verified by recombining modes  $\hat a_1$ and $\hat a_2$ on another beamsplitter and analyzing the state in the output. b) Implementation with a cloud of absorptive atoms. Detection of a re-emitted photon heralds a photon annihilation event and triggers recording of image on a CCD camera. Photon subtraction will not cast a shadow on the resulting quantum state, so its intensity distribution (solid orange line) does not change. This contrasts with regular linear absorption, which would cause a local shadow to appear in the intensity distribution (dashed blue line).}
\label{f1}
\end{figure}

Consider state $\ket\psi$ prepared in an optical mode defined by photon annihilation operator $ \hat a $; we assume all modes orthogonal to  $ \hat a $ to be in the vacuum state. This state is distributed, by means of a beam splitter, between remote parties Alice and Bob in modes $\hat a_1 $ and $\hat a_2$ such that $\hat a = \mu\hat a_1 + \lambda\hat a_2$ with $|\mu|^2$ and $|\lambda|^2$ being the nonvanishing beam splitter reflectivity and transmissivity, respectively [Fig.~\ref{f1}(a)]. Unless $\ket\psi$ is a coherent state or a statistical mixture thereof, this operation generates a state that is entangled with respect to the two parties \cite{Caves2013}.

Now suppose Alice applies the photon annihilation operator to her mode. We have
\begin{equation}
\label{eq1}
\hat a_1\ket\psi_{\hat a}=(\mu^*\hat a+\lambda\hat a_\perp)\ket\psi_{\hat a}=\mu^*\hat a\ket\psi_{\hat a},
\end{equation}
where $\hat a_\perp=\lambda^*\hat a_1-\mu^*\hat a_2$ is the annihlation operator of a mode orthogonal to $\hat a$. Because this mode is in the vacuum state, the action of its annihilation operator produces arithmetic zero. We see that the annihilation operator, albeit applied locally, acts upon the entire state $\ket\psi$ shared between the two parties.

Suppose, for illustration, that a cloud of weakly absorbing atoms is placed in a wide optical beam in mode $\hat a$ as shown in Fig.~\ref{f1}(b) \cite{Baragiola2012}. The mean probability of atoms to absorb a photon is much less than $1$. When an absorption event does occur, it is followed by reemission of a photon in a random direction, registered by a detector. A ``click" of this detector signifies application of photon annihilation to the mode $\hat a_1$ corresponding to the atomic cloud. One would expect the atoms to create a ``shadow" --- an area of reduced intensity --- in the laser beam. In fact, this does not happen; the intensity gets reduced uniformly over the entire laser profile, so mode $\hat a$ retains its structure. It is impossible to recover the position and shape of the atom cloud by looking at the output state of light. Hence the analogy with the folklore vampire that gave rise to the title of this paper.

The above argument may appear to contradict our everyday experience of observing shadows. The explanation is that shadows are caused by absorption of light, which is not equivalent to the annihilation operator. Rather, it is described by a Lindbladian $\partial\hat\rho/\partial z\propto\hat a_1\hat\rho\hat a_1^\dag-(\hat\rho\hat a_1\hat a_1^\dag+\hat a_1\hat a_1^\dag\hat\rho)/2$ (where $\hat\rho$ is the density operator of the state being attenuated and $z$ the direction of propagation) which contains both annihilation and creation operators. The latter, unlike the former, does not possess the nonlocal property described above: $a_1^\dagger \left|N\right\rangle_a \, \cancel{\propto} \left|N+1\right\rangle_a$ \cite{footnote}.

\begin{figure}[b]
\includegraphics[width=3.4in]{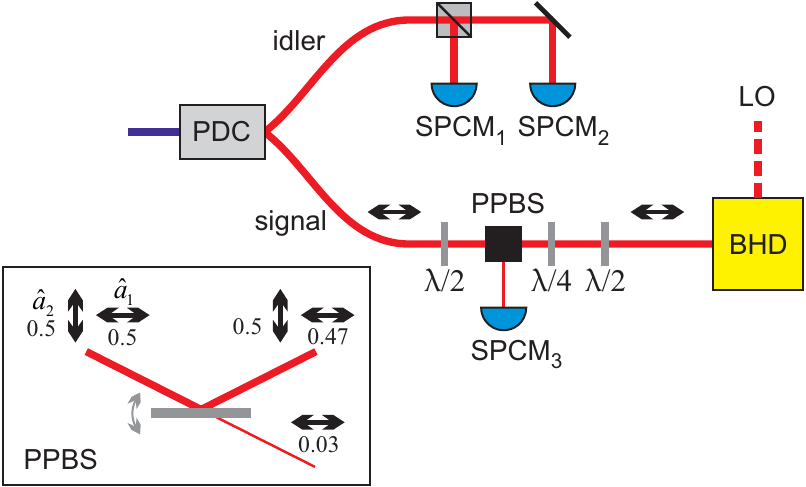}
\caption{Experimental setup. Mode $\hat a$ is prepared in the signal output of parametric down-conversion (PDC). The waveplates form a Mach-Zehnder  interferometer in the polarization basis, with modes $\hat a_1$ and $\hat a_2$ being its horizontally and vertically polarized arms. Photon is subtracted from mode $\hat a_1$ on the partially polarizing beam splitter (PPBS). Its improvised realization is shown in the inset, with arrows and numbers indicating polarizations of modes and their normalized intensities. The PPBS transmission, which is a compromise between the count rate and fidelity of subtraction, can be tuned by tilting the mirror.
BHD, balanced homodyne detection. LO, local oscillator. SPCM, single photon counting module.}
\label{p2}
\end{figure}

Importantly, the cloud of atoms in Fig.~\ref{p2}(b) must be weakly absorbing in order to implement the annihilation operator correctly. The cloud is not expected to affect the input state significantly or cast a shadow when it is being monitored without conditioning on the detector's click. However, when a click does occur, state $\ket\psi$ is known to have lost a photon. If it initially contains only a few photons, the relative loss of energy is significant. One would intuitively expect this loss to take the form of a shadow --- and yet it is not the case.

The action-at-a-distance of the photon annihilation operator can be made explicit by observing its effect on the mean number of photons in Bob's mode. If we start with Fock state $\ket N$ in mode $\hat a$, the photons are distributed between modes $\hat a_1$ and $\hat a_2$ in proportion with the beam splitter coefficient, so Bob's channel has $N|\lambda|^2$ photons on average. After Alice's application of $\hat a_1$, the state in mode $\hat a$ becomes $\ket{N-1}$, so the mean number of photons in Bob's mode changes, becoming $(N-1)|\lambda|^2$.

This observation does not imply superluminal signaling because photon annihilation is not a unitary operation, and as such can be realized only probabilistically. It is typically implemented by tapping a small portion of the target state onto a single-photon detector via a low-reflectivity beam splitter. A ``click" of the detector signifies a photon annihilation event \cite{Wenger2004,Ourjoumtsev2006,Ourjoumtsev2007,Takahashi2008,Ourjoumtsev2009,Kumar2012,Kurochkin2013}. One may argue that such a setting involves a measurement of the target mode and the nonlocal properties are thus not surprising. However, a fundamental difference between this implementation of the photon annihilation and regular von Neumann measurement is that in our case no collapse of the target state occurs.



\begin{figure}[t]
\includegraphics[width=\columnwidth]{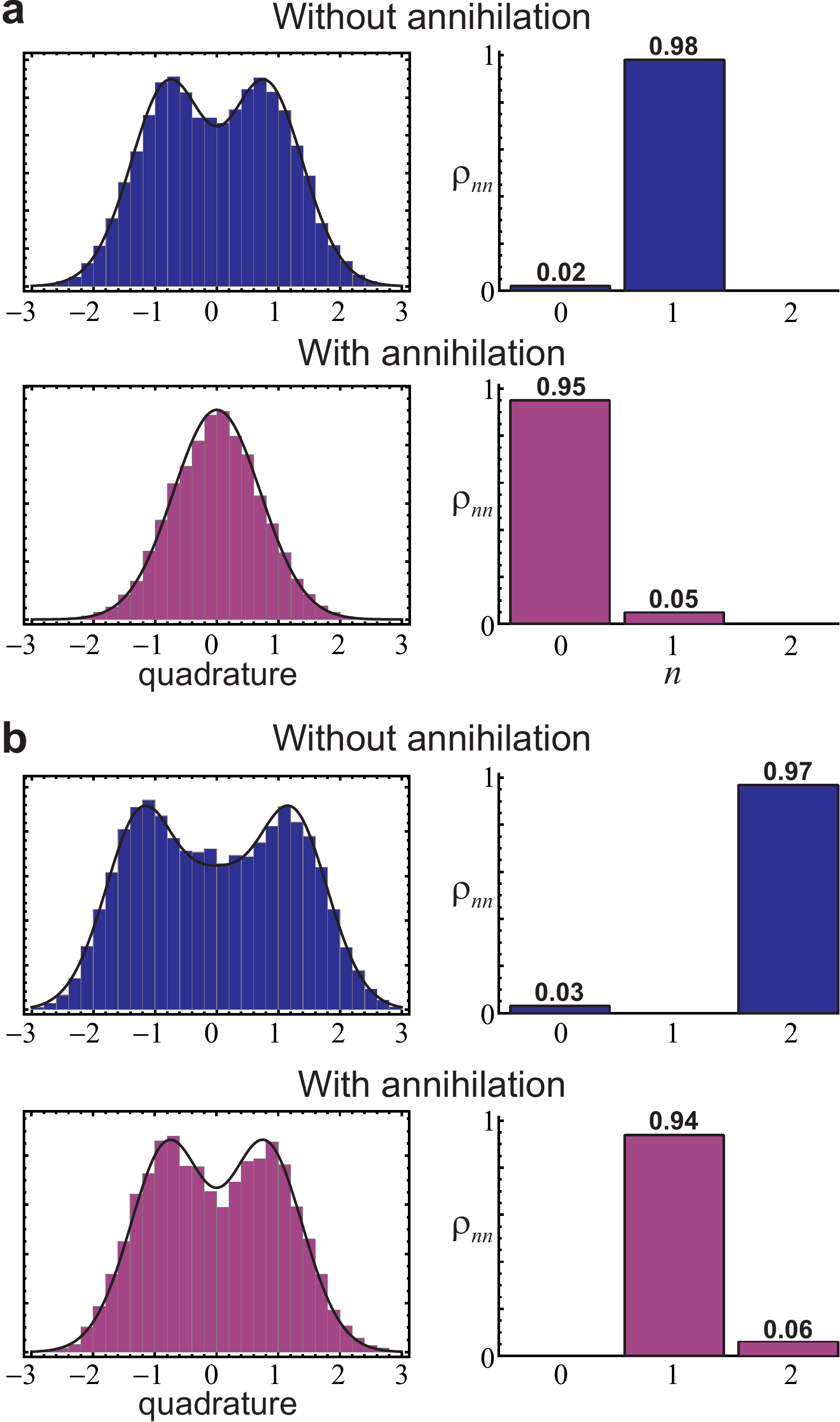}
\caption{Experimental results for the initial state of mode $\hat a$ being the one-photon (a) and two-photon (b) Fock states. The top row of each panel corresponds to the case without photon annihilation; the bottom row with photon annihilation applied to mode $\hat a_1$ (i.e. conditioned on SPCM$_3$ events). For each case, the quadrature histograms (marginal distributions) with theoretical expectation (lines) as well as the diagonal elements of the density matrix reconstructed with loss compensation are displayed.}  \label{p3}
\end{figure}

We demonstrate the quantum vampire effect experimentally, with modes $\hat a_1$ and $\hat a_2$ being horizontal and vertical polarization components of diagonally polarized mode $\hat a$. This corresponds to $\mu = \lambda = 1/\sqrt{2}$.

Mode $\hat a$ is initialized in a heralded Fock state. Type II parametric down-conversion takes place in a PPKTP crystal pumped at $390$nm with frequency-doubled pulses generated by a Ti:Sapphire laser with a repetition rate of 76 MHz and a pulse width of $\sim 1.8$ ps.
Clicks of one or two single-photon detectors [silicon-based single-photon counting module by Excelitas, SPCM$_1$ and SPCM$_2$ in Fig.~\ref{p2}(a)] in the idler channel herald synthesis of the one- or two-photon Fock state, respectively, in the horizontal polarization of the signal channel \cite{Ourjoumtsev2006a,Bimbard2010}. The heralded state is turned to a $45^\circ$ polarization using a $\lambda/2$ waveplate.

The annihilation operator is realized using a strongly imbalanced partially polarizing beam splitter. It employs a regular dielectric mirror coated for high reflectivity at a $45^\circ$ angle of incidence. The mirror is positioned to form an angle of incidence of $\sim 60^\circ$ with the incoming mode. The S polarization then still exhibits high reflection ($>99.8\%$) while in the P polarization about 6\% of the incident light is transmitted. The transmitted field is collected and detected using SPCM$_3$, so a click of that detector heralds with a high probability a photon annihilation event. Triple coincidences occur at a relatively low rate of $\sim 10$ counts per minute.

The polarization of the reflected light is then adjusted by a combination of a $\lambda/4$ and $\lambda/2$ plates, thereby returning mode $\hat a$ to horizontal polarization and completing the Mach-Zehnder interferometer in Fig.~\ref{f1}(a). The state of this mode is measured using homodyne tomography.

In the absence of a click from SPCM$_3$, this state is one- or two- photon Fock state, as evidenced by the top plots in Fig.~\ref{p3} (a,b). The overall efficiency of the reconstructed state is $\sim 53\%$. In addition to usual loss sources \cite{Lvovsky2001,Huisman2009}, about $1.5\%$ are due to the transmission of the partially polarizing beam splitter \cite{footnote1}. The diagonal elements of the corresponding density matrix in the photon number basis are reconstructed by the maximum-likelihood algorithm \cite{MaxLik04,MaxLik07} with loss compensation (Fig.~\ref{p3}, bottom row) to produce distributions that are consistent with the expected Fock states.

When SPCM$_3$ fires simultaneously with SPCM$_1$ and/or SPCM$_2$, photon subtraction from mode $\hat a_1$ occurs. As evidenced by the reconstruction results, the resulting state of mode $\hat a$ is the next lower Fock state, detected with the same efficiency. This indicates that the photon has been subtracted from the entire mode $\hat a$ without affecting its structure, as predicted theoretically. The remaining $6\%$ of the two-photon component in Fig.~\ref{p3}(b) are due to dark counts of SPCM$_3$.

The nonlocal property of the photon annihilation operator demonstrated here is of generic nature. It is expected to hold for optical modes in any basis (temporal, spatial, spectral, etc.), as well as for other bosonic systems. A related phenomenon, for instance, is known to occur in Bose-Einstein condensates: when a set of atoms is extracted locally from the condensate, the shape of the matter wave does not change.

We expect the quantum vampire effect to find applications in quantum information technology. For example, it enables non-local manipulation of quantum states without precise knowledge of their modes, such as in protocols for distillation of continuous-variable entanglement by photon annihilation \cite{Ourjoumtsev2007,Takahashi2010,Kurochkin2013}. The ability to ``steal" a photon without casting a shadow may prove useful for eavesdropping in quantum key distribution as well as developing quantum cloaking devices. We also believe the effect to be of fundamental interest, as quantum action at a distance that is not associated with a local state collapse has not yet been studied. 

\textit{Acknowledgment.} We thank Gora Shlyapnikov, Aleksei Fedorov, Yulia Shchadilova, Phillipe Grangier, Charles Bennett, Aephraim Steinberg and Thomas Jennewein for useful discussions. We thank Russian Quantum Center and NSERC for support. AL is a CIFAR Fellow.


\begin{thebibliography}{}
\bibitem{footnote} This is easily seen from \eeqref{eq1}: if $\hat a$ is replaced by $\hat a^\dag$, the term $\lambda\hat a_\perp^\dag\ket\psi_{\hat a}$ no longer vanishes.
\bibitem{footnote1} Mode $\hat a$ is equally split between the horizontal and vertical potatizations, so the 6\% transmissivity of the PPBS for the horizontal polarization implies a 3\% mean loss for mode $\hat a$. Because the efficiency of this mode is about $0.5$ due to other loss sources, the effect of the PPBS on that efficiency is $\sim 3\%\times 0.5= 1.5\%$.
\end{thebibliography}
\end{document}